\documentclass[12pt,preprint]{aastex}
\usepackage{verbatim}
  
\newcommand{\beq}{\begin{equation}}
\newcommand{\eeq}{\end {equation}}
 
\def\arcsecf {\hbox{$.\!\!^{\prime\prime}$}}

\begin{document}
\title{Four years of optical turbulence monitoring at the \\ Cerro Tololo Inter-American Observatory (CTIO)}
\author{S.G. Els\altaffilmark{1,2},  M. Sch\"ock\altaffilmark{3}, E. Bustos\altaffilmark{1}, J. Seguel\altaffilmark{1}, J. Vasquez\altaffilmark{1}, D. Walker\altaffilmark{1}, \\
R. Riddle\altaffilmark{2},W. Skidmore\altaffilmark{2}, T. Travouillon\altaffilmark{2}, K. Vogiatzis \altaffilmark{2}}
\altaffiltext{1}{Cerro Tololo Inter-American Observatory, National Optical Astronomy Observatory, Casilla 603, La Serena, Chile}
\altaffiltext{2}{TMT Observatory Corporation, 2632 E. Washington Blvd., Pasadena CA 91107, USA}
\altaffiltext{3}{TMT Observatory Corporation, 5071 West Saanich Road, Victoria, British Columbia, Canada }

\begin{abstract}
The optical turbulence conditions as measured between 2004 until end of 2008 above Cerro Tololo, their seasonal as 
well as nocturnal behavior are presented.
A comparison with the MASS-DIMM system of the Thirty Meter Telescope site testing was conducted and identifies 
an artificially increased seeing component in the data collected by the CTIO DIMM system under northerly winds. 
Evidence is shown that this increased turbulence is caused by the telescope dome. A correction for this effect is 
attempted and applied to the CTIO DIMM data. The MASS data of this comparison campaign allow to set constraints on 
the general assumption of uniform turbulent layers above a site. 
\end{abstract}
\section{Introduction}
CTIO on Cerro Tololo is one of the major U.S. observatories in the southern hemisphere. With 2200~m above sea level (a.s.l.) Tololo is a medium high mountain, being located approximately 
60~km east of the Pacific Ocean at the southern end of the Chilean Atacama desert. The site testing which led to the selection of this mountain (\citealt{stock64}), 
showed the outstanding quality of the astronomical observing conditions found at mountains in the Atacama desert. Even though not being the first investigation of its 
kind in this region (\citealt{curtis09}), it was in particular the site testing of Cerro Tololo which ultimately shifted the attention of other observatory projects 
to this part of South America (e.g., \citealt{blaauw91}). Also for the future generation of Extremely Large Telescopes the western 
coastal area of South America is one of the possible regions to host such a facility (e.g., \citealt{schoeck09},  \citealt{osip08}). Being thus one of the initial triggers 
for the astronomical development in Chile, the atmospheric seeing conditions at CTIO have been monitored over a long period of time. However, it turns out that the data 
which have been collected over the past decades were obtained with very heterogeneous instrumentation, making it difficult to put Cerro Tololo's observing conditions into 
perspective. In particular the atmsopheric turbulence distribution, which drives the seeing, requires appropriate instrumentation to be accurately measured. 
The Differential Image Motion Monitor (DIMM, \citealt{sarazin}) has become the prime tool to measure the total seeing.
But modern astronomical observing techniques also require detailed information on the vertical distribution of the optical turbulence strength, expressed by the 
refractive index structure constant $C_n^2$, at all heights $h$; the turbulence profile $C_n^2(h)$. Over recent years, the Multi Aperture Scintillation Sensor 
(MASS, \citealt{toko02}) has established itself as one of the main tools to obtain low resolution turbulence profiles. First experiments of MASS and DIMM measurements 
at CTIO were conducted during a two months long campaign and were published in \cite{tokovinin03}. This shed first light on the vertical atmospheric turbulence distribution 
above Cerro Tololo. However, campaigns conducted over relatively 
short periods of time always have the disadvantage of not being able to assess seasonal variability and therefore are likely to be biased in one way or the other. 
Long term monitoring of site conditions is important to assess turbulence variations on all time scales and to obtain the representative, or typical, observing conditions 
at a particular site. In April 2004 a combined MASS-DIMM instrument was thus permanently installed at CTIO. By now, more than four years of MASS and DIMM data have been 
obtained at Cerro Tololo and should allow a better view on the atmospheric conditions at this observatory. 

Between the years 2004 and 2005, CTIO was also hosting the experiments to calibrate the MASS-DIMM instruments of the site testing program for the Thirty Meter Telescope (TMT). 
As the TMT site monitor operated simultaneously with the CTIO site monitor, this allows for the calibration of the CTIO site monitor with respect to the TMT site monitoring equipment. 
As the data collected by the CTIO site monitor cover the same four years (2004--2008) as the TMT site monitoring program, this allows us to compare the 
turbulence conditions encountered above the southern Atacama ($\sim 30^\circ$S) and the central Atacama ($\sim 24^\circ$S) desert, where three of the TMT candidate sites are 
located. 

In this article, we describe the site monitoring system at CTIO (\S~\ref{ctiomonitor}), compare its MASS-DIMM data to the TMT site testing instrumentation 
(\S~\ref{ctiotmtcomp}) and provide the results of the four years of atmospheric turbulence observations which have been collected on Tololo (\S~\ref{20042008}). 

\section{The CTIO site monitoring system}
\label{ctiomonitor}
The CTIO site monitor consists of a Meade LX-200 telescope with an aperture of 25~cm, mounted on its standard alt-az fork mount. The telescope is housed on top a 6~m tall 
tower and is protected by a vinyl dome. The setup is shown in Fig.~\ref{ctiodimmtower}. During the year 2000 a DIMM (\citealt{boccas01}, \citealt{tokovinin03}) was installed 
at this telescope and in early 2004 replaced by a combined MASS-DIMM unit (\citealt{kornilov07}). Here we will focus on the MASS-DIMM data only, collected since March 2004. 

The DIMM measures the total seeing, which is inverse proportional to the Fried parameter $r_0$, by means of the variance of the image separation of a stellar image observed 
through two subapertures. The specifications of a MASS-DIMM device in combination with the LX-200 telescope and a ST-5 CCD are provided in \cite{kornilov07}. In brief, this 
results in two (slightly vignetted) subapertures with a diameter of 8~cm, separated by 17~cm. The CTIO DIMM software was modified over the years but the main code remained the 
same to what is described in \cite{tokovinin03};  consecutive frames are taken with 5~ms and 10~ms exposure time and then corrected to zero exposure time by means of a two point 
exponential extrapolation (\citealt{tokovinin02a}). Each measurement sequence takes 60~s. As the readout speed depends on the employed computer architecture, this does 
not result always in the same amount of subframes. Typically each DIMM measurement consists of more than 180 subframes for each exposure time. Including processing overhead, 
a DIMM measurement is triggered approximately every 70~s. 

MASS (\citealt{toko02}) reconstructs, by measuring the differential scintillation indices in four concentric subapertures, 
a low resolution vertical profile of the optical turbulence structure constant or turbulence profile: $C_n^2(h)dh$ at altitudes $h_i = 0.5, 1, 2, 4, 8, 16$~km. 
The $C_n^2(h)dh$ value provided by the MASS is the structure constant integrated over the altitude range $dh$, weighted by a weighting function. 
The weighting function for each layer peaks at its nominal altitude $h_i$ and drops to zero at the altitude of its adjacent layers, thus causing some overlap 
between neighboring layers. 
From these turbulence profiles the isoplanatic angle $\theta_0 \propto (\sum_{i=1...6} C_n^2(h_i) h_i^{5/3}dh)^{-3/5}$ and the seeing $\epsilon_{\mathrm{MASS}} \propto (\sum_{i=1...6} C_n^2(h_i))^{3/5} $, 
which would be seen by an observer 500~m above the telescope, can be inferred. The MASS also provides, under certain assumptions, some measurement of the atmospheric coherence time $\tau_0$ 
(\citealt{toko02}). These, however, require some specific calibration factor for which we use the value provided in \cite{travouillon09}.
The MASS is operated by the {\tt turbina} software (\citealt{kornilov07}, \citealt{toko07}) and measurements are triggered simultaneously with the DIMM measurements. 
The flux in each subaperture is measured by a photomultiplier tube, collecting 1~ms samples over a total of 60~s. 

In this article, the turbulence parameters, apart from $\tau_0$, have been corrected for their zenith distance and are given at a wavelength of $\lambda=500~$nm. 
$\tau_0$ shows a very complex zenith distance dependence and the induced error by not correcting to zenith is expected to be small as outlined in \cite{travouillon09}.

The CTIO site monitoring equipment also consists of commercialy available meteorological sensors by R.M. Young. Wind speed and direction are measured, as well as air 
temperature, relative humidity and atmospheric pressure. These sensors are installed on a 30~m tall tower, which is located on the steep eastern slope 
of Cerro Tololo. Therefore, the sensors are ranging effectively up to approximately 6~m above the Tololo summit platform. Data were acquired every 2 to 5~min. 
No particular data quality monitoring of these data was in place and therefore the measurement accuracy and precision of the sensors is not known and might be 
not as good as what is reported for other site monitoring projects (e.g., \citealt{riddle09}). 
\section{The CTIO versus the T3 MASS-DIMM}
\label{ctiotmtcomp}
The TMT site testing program operated between August and October of 2004, two of its site testing telescopes (refered to as T2 and T3) side by side on Cerro Tololo. 
This campaign was conducted in order to indentify the precision of the MASS and DIMM data which would be collected during the TMT site testing. 
Results of this particular campaign were reported by \cite{wang07} and \cite{els08} for DIMM and MASS, respectively. This campaign was terminated in October 2004 by 
the move of T2 to a TMT candidate site in northern Chile. However, the T3 site testing telescope remained on Tololo until October 2005. Even though its operational scheme 
was during this remaining time not kept as efficient as during the previous calibration campgain, it still resulted in a significant amount of MASS-DIMM data collected 
in parallel with the CTIO site monitor, covering the time span between October 2004 and October 2005. These were continous observations but partly with large gaps in time, 
in particular during the second half of the campaign (June -- October 2005). 
We use these data in order to compare the CTIO MASS-DIMM results to the T3 results and therefore to the results obtained at the TMT candidate sites. 
\subsection{MASS data}
\label{masscomp}
The MASS instruments at T3 and CTIO are of the same design, as described in \cite{kornilov07}. However, they are mounted on different telescopes 
and the light paths thus contain different optical elements, affecting the calibration of these systems. The instrumental parameters, like for example 
the spectral response and instrument magnification, are defining the spatial filter of the scintillation pattern. 
Knowing these parameters with good accuracy is therefore essential for a proper turbulence profile reconstruction using differential scintillations. 
Some of these parameters can only be obtained reliably, after the observation. Therefore, 
the data sets were reprocessed using the {\tt atmos} package within the  {\tt turbina 2.052} software (\citealt{kornilov07}, \citealt{toko07}), employing 
the appropriate settings for each telescope instrument combination at the time of observation. Only those data from T3 and CTIO were selected which 
were taken within 60~s of each other. This resulted in 41,128 simultaneous T3--CTIO pairs of MASS measurements, including observations when the systems 
were not pointing at the same star. 
28,029 measurement pairs were obtained when both systems observed the same star. The comparison of the turbulence statistics for observations 
when T3 and CTIO were pointing at the same star is shown in Tab.~\ref{ctiot3masscomptab}. The difference between the observed median seeing values 
is less than 0\arcsecf02 and thus even lower than the 0\arcsecf05 which which were found during the TMT campaign for the precision of two idential 
MASS systems (\citealt{els08}).  
On the other hand, the scatter ($rms$ in Tab.~\ref{ctiot3masscomptab}) in the correlation between the CTIO and TMT MASS is a factor of two larger than 
what was found for the TMT MASS.   
This is an indication of the measurement uncertainty and might also be a affected by the 60~s selection criteria employed here and the slightly larger 
spatial separation of the telescopes of 10~m; for the TMT study it was 30~s and 5~m, respectively. 

\begin{figure}[h]
\begin{center}
\resizebox{0.75 \textwidth}{!}{\rotatebox{90}{\includegraphics{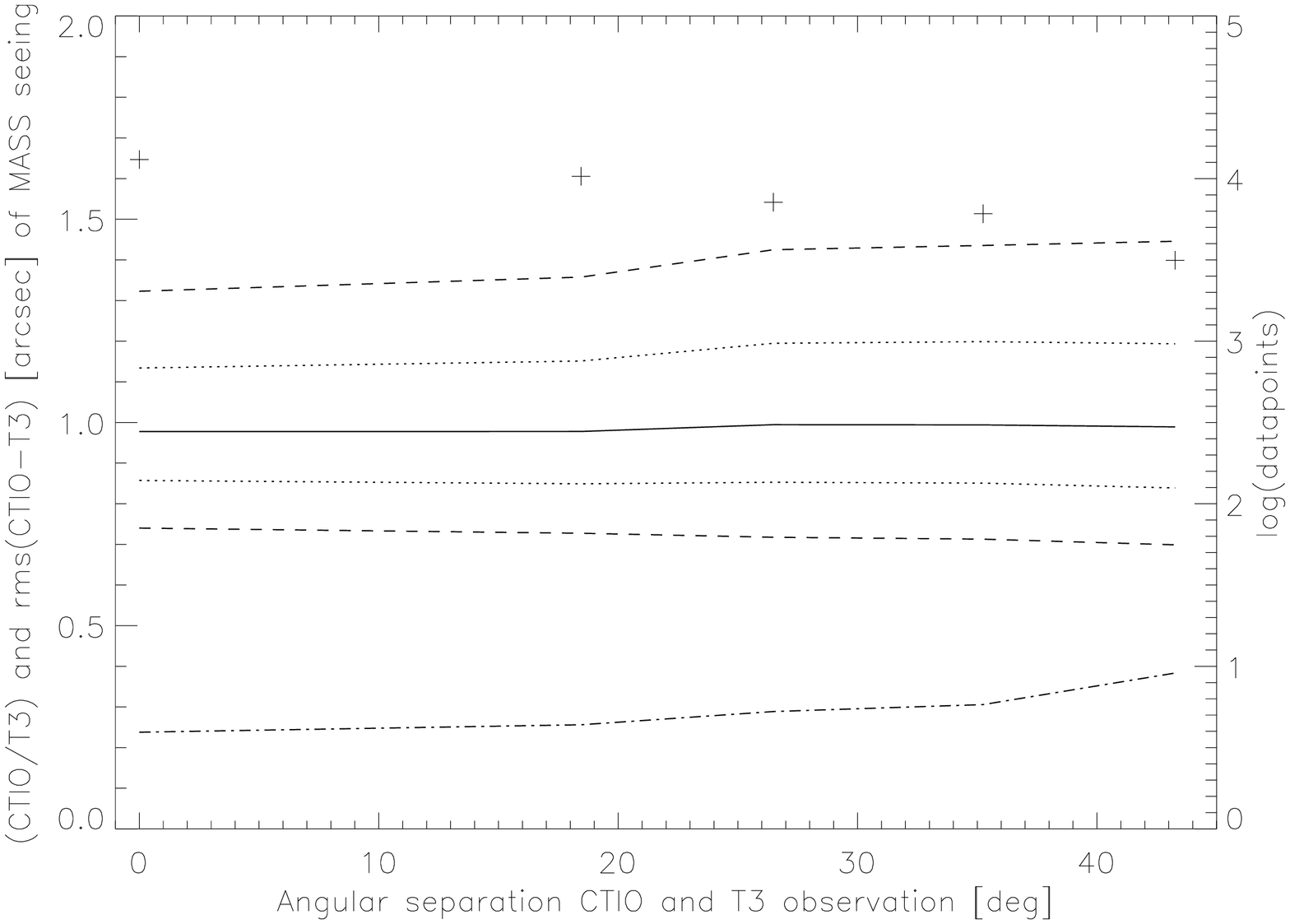}}}
\resizebox{0.75 \textwidth}{!}{\rotatebox{90}{\includegraphics{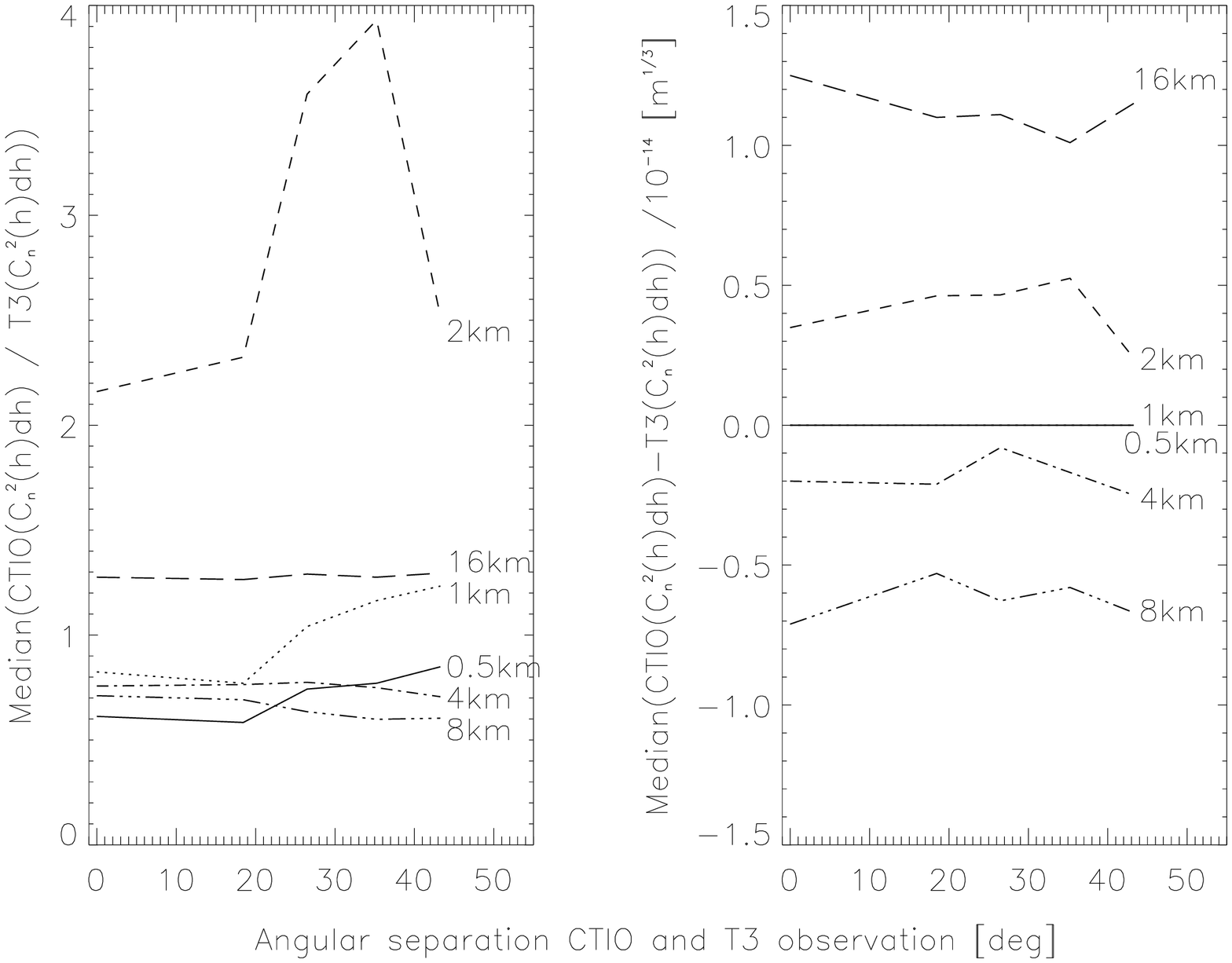}}}
\end{center}
\caption{\emph{Top panel:} Statistics of the MASS seeing ratio between CTIO and T3 observation pairs. The dashed lines show the 10\% and 90\%iles, 
the dotted lines the 25\% and 75\%iles, the solid line the median at each separation, respectively. The dashed-dotted line is the $rms$ of 
the seeing data at each separation. Crosses represent the amount of data at each separation. 
\emph{Lower panel:} Median of the difference between CTIO and T3 turbulence strengths of each MASS layer. Layers indicated by the same line styles 
as in the other panels. Note, the results for the 0.5 and 1~km are basically zero and are indicated by the solid line. 
  }
\label{angularmasssep}
\end{figure}

Figure~\ref{angularmasssep} addresses the effect of turbulence in different directions in the sky. As the data which are covering the various separations 
are spread out over the entire duration of the campaign, they do not represent a peculiar weather pattern or observation configuration.
The top panel shows the statistics of the ratio of the MASS seeing pairs for the different angular separations (this is different to what is shown in 
Tab.~\ref{ctiot3masscomptab} where the difference between the statistical distributions are shown). The almost 
perfect median agreement over the entire range of separations demonstrates the average 
isotropic behavior of the free atmosphere seeing. The mild increase of the $rms$ with increasing angular separation (approximately by 0\arcsecf1 over 40$^\circ$) 
and that the differences between the 90 percentile and the 10 percentile remain on an almost constant level, demonstrate 
that the free atmosphere seeing does on average not differ more than approximately $\pm30\%$ or $\pm0\arcsecf1$ between different directions in the sky. 
Our analysis does not separate between different azimuth and altitude directions in the sky, e.g, here we do not map the turbulence across the sky as was 
done in the simulations shown in \cite{masciadri02}. Our observed simultaneous differences in free atmosphere seeing (up to $rms = 0\arcsecf38$ at 43$^\circ$) 
between different directions in the sky are comparable to the $0\arcsecf2$ which were observed by \cite{masciadri02} during a single night. A quantitative 
comparison with the results of this reference is not possible as they apply to different sites with different topographic features. 

The individual layer strengths behavior is more complex and is shown in the lower two panels of Fig.~\ref{angularmasssep}. 
The medians of the quotients of the layer strengths are shown in the lower left panel of that Figure. The relative differences of the upper three layers remain almost 
constant over the entire range of separations. In comparison the layers up to 2~km show an increase of their relative difference beyond 20$^\circ$ separation. 
The lower right panel of Fig.~\ref{angularmasssep} shows the median difference between the layer strength as measured by the two systems. 
This difference is zero for the 0.5~km and 1~km layers and is explained as these layers are basically turbulence free. 
The higher layers show stronger turbulence and the absolute differences can reach $1.3\cdot 10^{-14}$~m$^{1/3}$ (equivalent to 0\arcsecf093) in the 16~km layer. 
As these absolute differences remain almost constant over the entire range of observed separations, this indicates systematic differences 
of the reconstruction of the individual layer strengths by the two MASS systems.  The variation around the mean of each of these curves indicates that the assumption 
of turbulence above the site being on average uniformily distributed in vertical, parallel layers is valid to within approximately $0.2\cdot 10^{-14}$~m$^{1/3}$ 
(equivalent to 0\arcsecf030). 
\subsection{DIMM data}
Comparing the DIMM data obtained by these two site monitoring systems is more complicated than the MASS data. The DIMM channels of these systems 
differ not only in the employed hardware but also in the used analysis algorithms, despite their being based on the same principles. The T3 system uses a ST-7 CCD 
in scanning mode which provides a binned one dimensional frame of a fixed area on the CCD, with exposure times of 6~ms during a 36~s measurement cycle. 
In comparison the CTIO DIMM makes use of a ST-5 CCD which delivers images of 100~px $\times$ 17~px with exposure times of 5~ms and 10~ms over 60~s. 
The T3 scanning mode results in significantly more images during a measurement cycle than the two dimensional imaging technique; 6000 versus approximately 180. 
But T3 thus obtains only a 36~s average of the seeing as compared to the 60~s average obtained by the CTIO DIMM. 
For each measurement cycle of the CTIO DIMM, the area read by the CCD is adjusted around the locations of the stellar images; in turn T3 does the centering of the images 
inside the measurement area, by offsetting the telescope after taking a pointing image. The centroiding of the stellar images to determine the differential motion is 
done in the CTIO DIMM by calculating the center of gravity, whereas in the T3 DIMM two Gaussians are fitted to the one dimensional stellar images. The extrapolation to zero 
exposure time is realized in the T3 system by means of rebinning consecutive images to either six or two different exposure times. The CTIO DIMM in comparision 
performs the extrapolation based only on two exposure times only. And finally, while the T3 DIMM only measures the longitudinal (parallel to the subapertures) seeing 
component, the CTIO DIMM provides both components and the average of the two. 
Despite all these differences, we expect that both systems provide an accurate measure of the atmospheric turbulence. 

The DIMM data sample of T3 covers only the time up to May 2005. After that date the T3 telescope showed some misalignment, resulting in low Strehl values 
of the DIMM images, thus compromising the seeing results (\citealt{wang07}). Similar to the MASS data, only DIMM data taken within 60~s by both systems are considered 
in the present study, resulting in 39,154 samples, including observations of different stars. 
From these T3 DIMM data we find a median difference of 0\arcsecf011 between the seeing obtained by the six point extrapolation scheme and the formula by \cite{toko02} 
for two exposure times. 
This value is less than the precision of TMT DIMM data of 0\arcsecf02 (\citealt{wang07}). And as the reported results from the TMT candidate sites (\citealt{schoeck09}) 
refer to the six point extrapolated values, we will in the following make use of these DIMM seeing values from T3. 
The median difference between the longitudinal and transversal seeing components recorded by the CTIO DIMM is 0\arcsecf069. We consider this difference to be too large 
in order to make use of the average of these components for comparison purposes. Instead, we will use the longitudinal seeing component provided by the CTIO DIMM. Apart from 
T3 DIMM measuring this particular seeing component only, the longitudinal component has also the advantage that it is less sensitive to various errors. On the one 
hand the constant $K_l$, which relates the differential image motion to the Fried parameter $r_0$ and thus to seeing, is about 50\% smaller than the transversal 
constant $K_t$ (see eq. [8] in \citealt{toko02}). Also the longitudinal seeing appears less sensitive to the orientation between the wind direction and the 
axis of the DIMM subapertures (\citealt{toko02}). 

\subsubsection{Discussion of CTIO DIMM data prior to July 12 2005}
\label{biassec}
To complicate the comparison between these DIMM systems further, the CTIO DIMM underwent a correction of its analysis software, which took place in July 2005, 
shortly after the collection of useful T3 DIMM data was terminated on Tololo. Therefore, we will first investigate the impact of this modification and correct 
the CTIO DIMM data taken prior to this software change. Prior to the modification, the CTIO DIMM software did perform the centroiding on images which 
contained a digital bias, introduced by the ST-5 controller to avoid negative pixel values. In addition, the centroiding window around an initially identified stellar 
image location, was set to a radius of $r=6$~pixel. 
After July 12, 2005 the code was changed in such a way that the median background value $b$ is computed from the area outside the window regions, 
with radii of $r=4$~pixel. This background is subtracted from the entire image and final centroiding 
takes place inside the 4~pixel radius window, in accordance with the recommendations given in \cite{toko02}. 

\begin{figure}[!h]
\begin{center}
\resizebox{0.8 \textwidth}{!}{\rotatebox{90}{\includegraphics{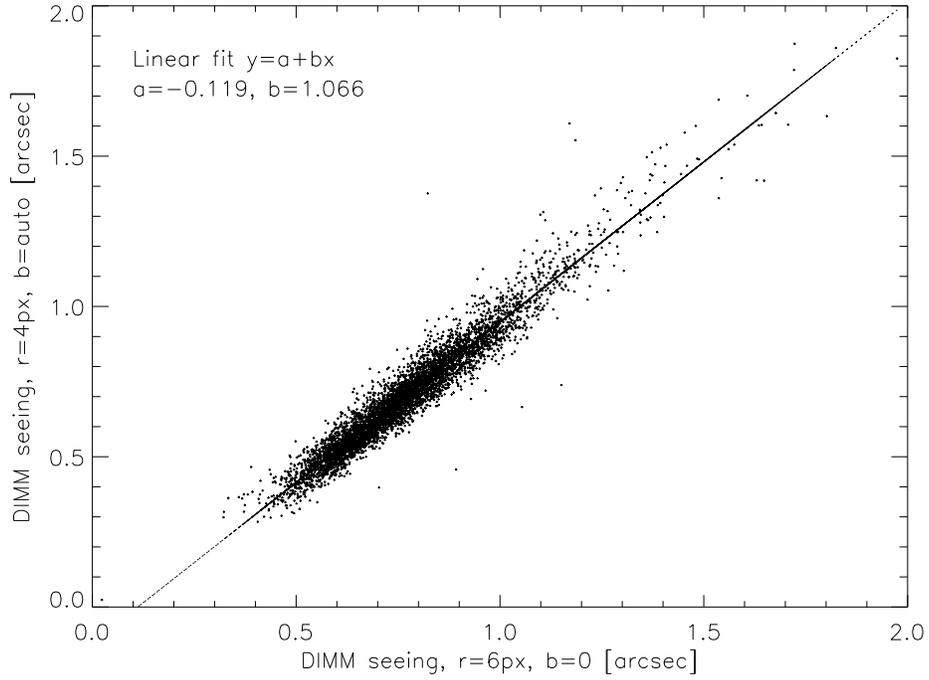}}}
\end{center}
\caption{CTIO DIMM data observed during April 2009 and reanalysed in two ways. On the x-axis the the DIMM analysis as performed before July 2005 
is shown and on the y-axis the results according to the current analysis method is indicated (see text for details). The line shows a linear fit 
to the data, whose parameters are shown as well. }
\label{backgroundtest}
\end{figure}

Obviously, the subtraction of a constant background and using a different window size affects the centroid determination. 
To assess from an empirical point of view the impact of these changes, a stand-alone DIMM reanalysis package was written using the existing CTIO DIMM routines. 
A total of 5465 DIMM observations were taken by the CTIO DIMM between April 2 and April 20, 2009 and the raw, two dimensional DIMM frames saved. These frames were 
analysed employing the software configurations before and after July 2005, meaning $r=6~$px, $b=0$ and $r=4~$px and $b$ being determined automatically. The correlation 
between the so obtained DIMM longitudinal seeing measurements is shown in Fig.~\ref{backgroundtest}. The configuration with the wider window size and no background subtraction 
results in larger seeing as compared to the currently employed configuration. We note, that for the transversal seeing component a very similar correlation and fit coefficients are found. 
We use the fit coefficients provided in Fig.~\ref{backgroundtest} and apply this fit to the CTIO DIMM data obtained before July 12 2005. 
\subsubsection{The enhanced ground layer at Tololo}
\label{glsec}
After applying the correction developed in the previous section to the CTIO DIMM data, the comparison between the distributions of the CTIO and T3 DIMM data 
shows a difference of the median seeing of 0\arcsecf082, which is still a factor of four larger than the DIMM seeing precision reported by \cite{wang07} for the 
TMT DIMM seeing monitors. The difference does not change if one considers only observations for which both systems were observing the same star\footnote{Note that  
the longitudinal axes of the subapertures of both DIMM systems are parallel to the elevation axes of the telescopes, thus, both DIMM systems experience the same 
attenuation of the DIMM response due to the alignment of wind direction with the subaperture axis if pointing at the same star.}. After the encouraging results 
from the MASS section this finding requires some more investigation. 

\begin{figure}[h]
\begin{center}
\resizebox{0.8 \textwidth}{!}{\rotatebox{90}{\includegraphics{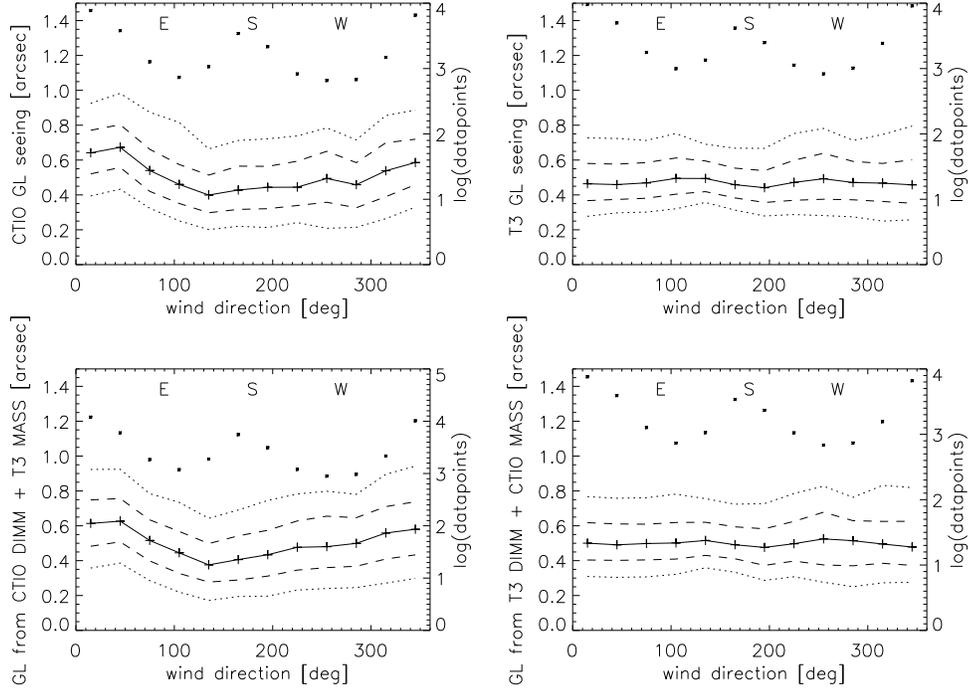}}}
\end{center}
\caption{Ground layer (GL) seeing roses from simultaneous observations by CTIO and T3 MASS-DIMM. The solid lines and the 
crosses indicate the median seeing values in each 30$^\circ$ wind direction bin, the dotted lines the 10\% and 90\%iles and 
the dashed lines the 25\% and 75\%iles. Dots mark the amount of data in each bin.
\emph{Top left:} CTIO MASS-DIMM GL;
\emph{Top right:} T3 MASS-DIMM GL;
\emph{Lower left:} CTIO-DIMM and T3-MASS computed GL;
\emph{Lower right:} T3-DIMM and CTIO-MASS computed GL.
 }
\label{ctiot3GLroses}
\end{figure}

In a previous study of the conditions at Cerro Tololo, \cite{tokovinin03} found that the ground layer (GL) seeing, computed from the 
difference of MASS and DIMM seeing, increases when winds are coming from northern directions. This is at first counter-intuitive as the CTIO site monitor 
is located at the northern most edge of the Tololo summit platform. It would rather be  expected that southern winds cause the air to become more 
turbulent when passing over the summit area with its various buildings. Using the simultaneous T3 and CTIO seeing data, as well as simultaneously 
(within 120~s) recorded wind direction measurements, we construct what we call ``seeing roses''. The wind directions are binned 
in $30^\circ$ bins and the GL seeing statistics are calculated for data within each wind direction bin. An increase under northern winds is clearly visible in the 
CTIO GL seeing rose, which is shown in the top left panel of Fig.~\ref{ctiot3GLroses}. On the other hand, the T3 GL seeing remains at an almost constant level 
independent of the wind direction (top right panel in Fig.~\ref{ctiot3GLroses}). In order to test whether the observed differences are caused by the MASS or DIMM data, 
we computed GL seeing roses (lower two panels of Fig.~\ref{ctiot3GLroses}) using the T3 MASS data in combination with the CTIO DIMM data and the T3 DIMM data in combination 
with the CTIO MASS data. The strong increase of GL seeing is present only when the CTIO DIMM data are used. This clearly demonstrates, that the CTIO-DIMM is showing a 
seeing bias under northerly winds. The CTIO GL seeing roses show that this bias is strongest for winds between 300$^\circ$ and 120$^\circ$, thus wind directions centered 
around approximately 30$^\circ$ (North-North-East). 

\begin{figure}[h]
\begin{center}
\resizebox{0.8 \textwidth}{!}{\rotatebox{0}{\includegraphics{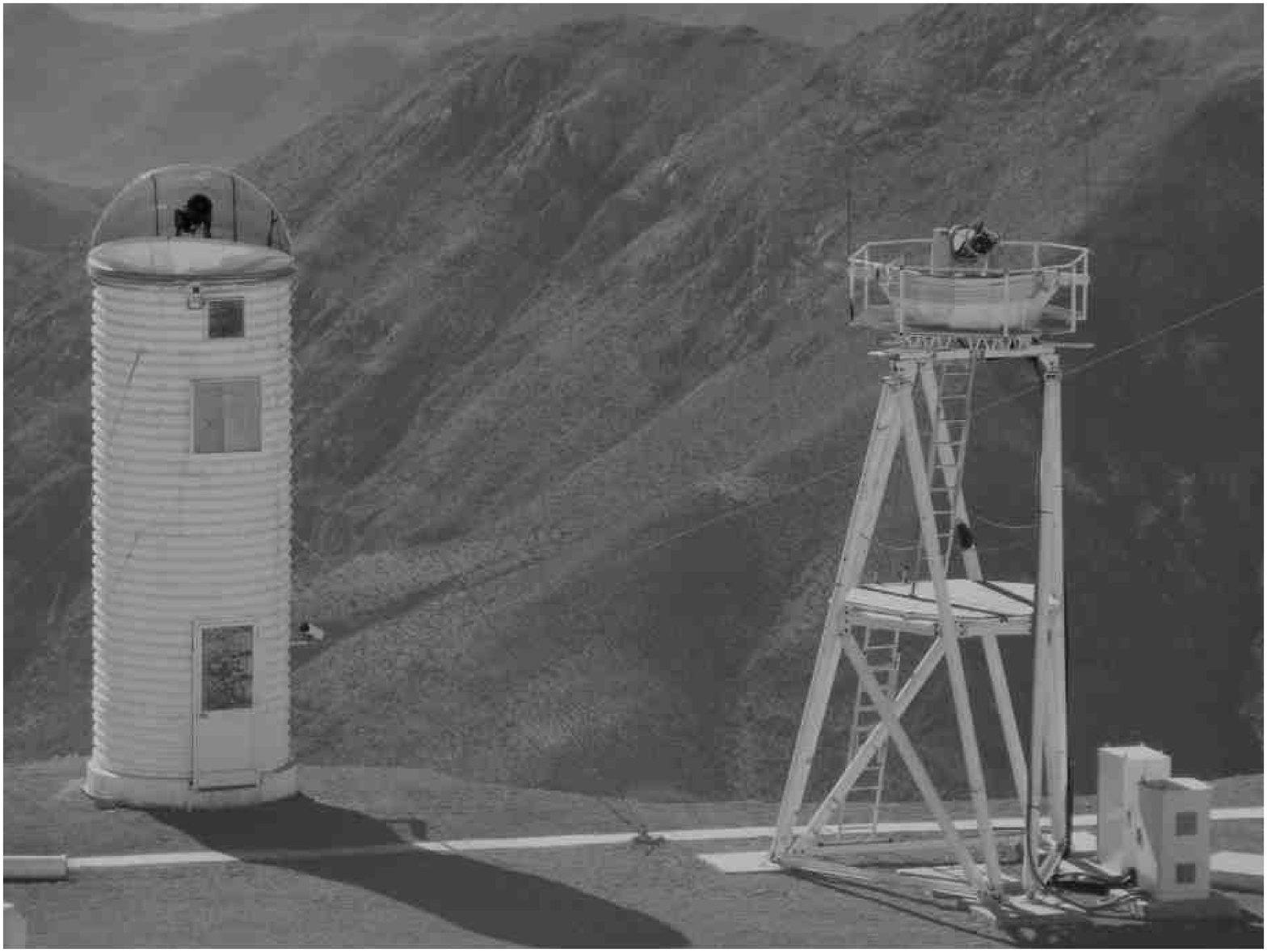}}}
\end{center}
\caption{Photograph of the CTIO site monitor (white tower to the left) and the T3 system (tower on the right) at the 
northern edge of the summit platform of Cerro Tololo. The East-West direction is indicated by the white conduit covers on the ground. 
This image shows the present (since April 2009) configuration. During 2004 and 2005 the T3 system was located 1~m SE of its current 
location. Both telescopes are mounted on 6~m tall towers. The image shows the systems with their domes open. 
 }
\label{ctiodimmtower}
\end{figure}

The clear dependence of this GL bias on the wind direction indicates that it is not an inherent problem of the CTIO-DIMM software 
and that the cause must be located within the close vicinity of the CTIO site monitoring telescope. Comparing the 
structural differences of the setup of the two site monitoring systems sheds light on this issue. Figure~\ref{ctiodimmtower} shows a picture of
the CTIO site monitor in its operational configuration. The telescope is mounted on a 6~m tall, approximately 50~cm$\times$50~cm concrete pier, 
which is surrounded by a metal shelter with a diameter of 2.6~m. The shelter tube has several openings at different altitudes above the ground, 
which are facing North-North-East, South and South-West. On top of the metal shelter a vinyl dome is installed, which opens 
towards south during observations but only by an angle of approximately 110$^\circ$. This means that during observations 
the dome remains closed up to an elevation of approximately 70$^\circ$ above the Northern horizon. We note, that the dome opens not exactly North-South but rather 
slightly towards South-West (maybe by 30$^\circ$). This only half opening dome is intended to prevent wind inducing vibrations into the telescope. 

Figure~\ref{ctiodimmtower} also shows the T3 setup on Tololo at its current (since March 2009) location, which differs from its location during the 2004--2005 
campaign by only 1~m (horizontally). The picture clearly shows the skeleton design of the tower hosting the 
T3 telescope. It also shows the dome open, as during night time operations. This demonstrates that the T3 dome fully folds away from the telescope during night time. 
The TMT site testing telescopes are custom built, open tube, 35~cm aperture Cassegrain telescopes. The entire design of 
telescope and tower are intended to minimize the influence of the telescope support structures on the airflow, as well as to provide maximum 
mechanical robustness. 

Here we cannot rule out that the various openings in the shelter structure of the CTIO site monitor are causing a vertical ``chimney-like'' 
flow through the shelter tube during northern winds and are thus thermally increasing the turbulence at the telescope. But due to its proximity and excellent alignment agreement 
of the seeing bias with the half-sphere of the dome, we suspect the partial lowering the dome of the CTIO telescope as the main source of the increased 
GL seeing observed by this telescope. Another argument supporting this hypothesis is illustrated in Fig.~\ref{ctiot3GLwsdiff}, where the dependence of the 
median ratio between GL turbulence strength as measured by the two site monitors on the wind speed is shown; separated are the cases of northern and southern winds. 
Based on the GL seeing roses in Fig.~\ref{ctiot3GLroses}, we define northern wind directions to be between 300$^\circ$ and 120$^\circ$ and southern winds for 
all other directions. In the case of southern winds, the quotient remaines almost constant between 0.8 and 1 over the entire range of wind speeds for which a meaningful 
amount of data has been collected. Interestingly, in the case of northern winds the quotient first raises from close to 1 up to 2 and above 3~m~s$^{-1}$ begins to decrease again. 
This decrease might hint on the influence of the geometry of the dome on the separation of the airflow and the behavior of the drag coefficient $c_v$ of the dome. 
With a diameter of the dome of $D=2.6$~m and the cylinder of the shelter building below, and assuming the kinematic viscosity of air being 
$\eta = 1.7147\cdot 10^{-5}$~m$^2$s$^{-1}$ (\citealt{lide95}), the Reynolds number $Re = ws D / \eta$ of the airflow around the CTIO site monitor dome is in the range  
of $10^5 \lesssim Re \lesssim 10^6$ for wind speeds between $ws=1$~m~s$^{-1}$ to 6~m~s$^{-1}$ (see upper x-axis of Fig.~\ref{ctiot3GLwsdiff}). 
For such a sphere the transition to a turbulent flow and the formation of a wake is expected to start at $Re=3\cdot 10^5$. At this point $c_v$ will decrease strongly. 
The wake dissipates and $c_v$ increases again beyond $Re=5\cdot 10^5$. This behavior of a flow around an object is well known (e.g,\citealt{brennen05}). 
Therefore, we speculate that our observations trace a connection between $C_n^2$ and the drag coefficient. Further measurements are needed to confirm and 
properly quantify such a connection. 
\subsubsection{Correction of the enhanced GL}
\label{corrsection}
\begin{figure}[h]
\begin{center}
\resizebox{0.8 \textwidth}{!}{\rotatebox{90}{\includegraphics{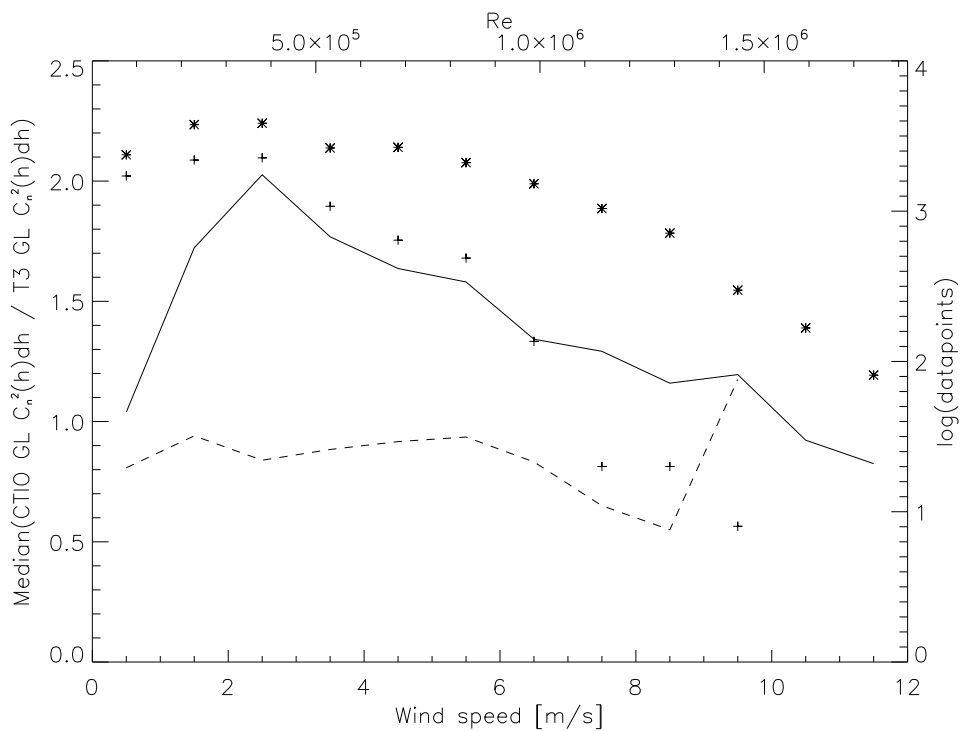}}}
\end{center}
\caption{Median difference between CTIO and T3 observed GL seeing measurements at various wind speeds. 
The top x-axis indicates the corresponding Reynolds number $Re$ of a shpere of the same diameter as the dome 
of the CTIO site monitor. Wind speeds were binned in 1~m~s$^{-1}$ intervalls. The solid line was computed from 
observations taken during northern winds and the dashed line was computed for southern winds. Crosses mark the 
amount of measurements during southern winds in each bin and asterisks for nothern winds, respectively. The total 
amount of these simultaneous MASS-DIMM data from T3 and CTIO and weather station data is 29725.
 }
\label{ctiot3GLwsdiff}
\end{figure}

\begin{figure}[h]
\begin{center}
\resizebox{1 \textwidth}{!}{\rotatebox{90}{\includegraphics{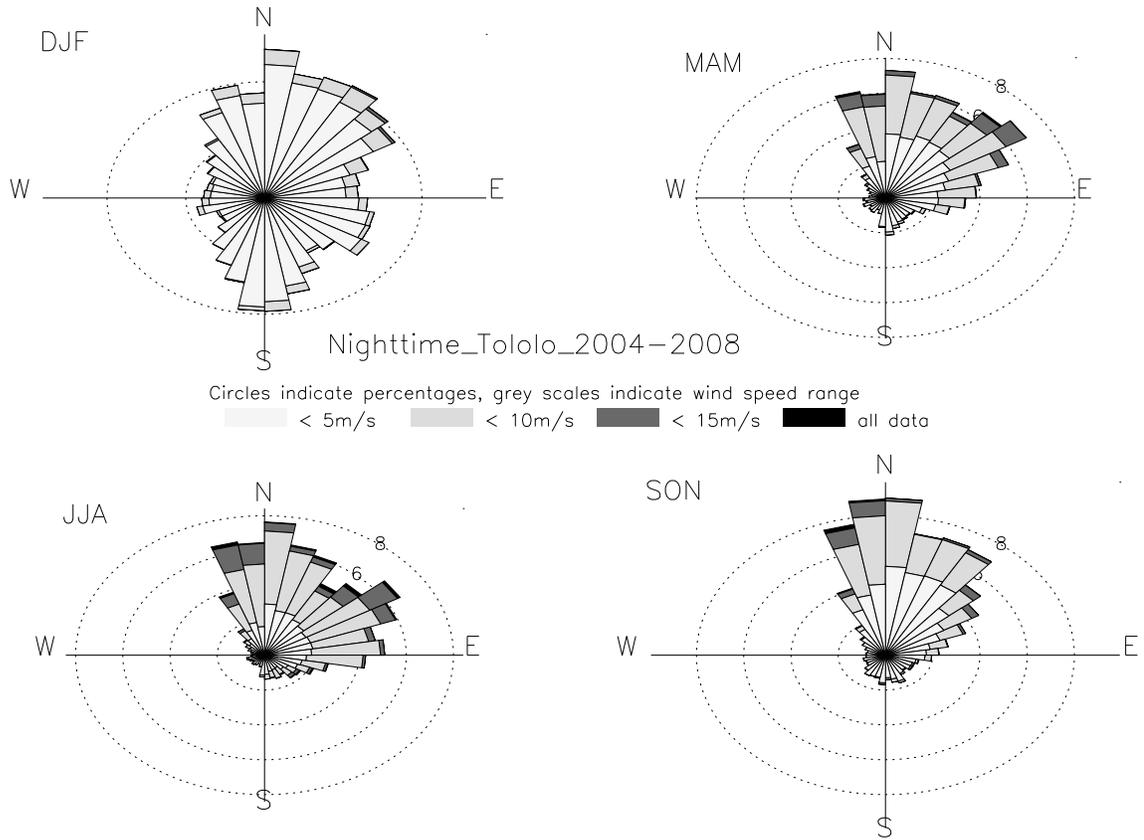}}}
\end{center}
\caption{Seasonal wind roses observed between January 2004 and December 2008. From top left to bottom right 
the wind roses during the summer, autumn, winter and spring months are shown. Circles are spaced by 2\% increments 
and indicate probability. Grey scale bars show the respecitve histograms for different wind speed ranges. 
The  wind roses were calculated from data measured during night time only, defined as times when sun elevation $< -12^\circ$. 
 }
\label{windroses}
\end{figure}

Even though we are aware of the uncertainties of trying to remove a dome/tower seeing effect from our measurements, we will attempt to do so as otherwise four years of 
DIMM data collected under northern winds at CTIO would become useless. And northern winds are dominating the wind rose of Cerro Tololo (see Fig.~\ref{windroses}).
The bias is localized in the GL and we can thus use the MASS-DIMM computed GL data, the MASS seeing and a corrective term for the GL measurement to obtain a corrected total 
seeing value in the following way:
\begin{equation}
	\epsilon_{\mathrm{DIMM,corr}} \propto \Big( \frac{1}{f_c(ws,wd)} C_n^2(GL)dh_{\mathrm{GL}} + \sum_{i=1,6} C_n^2(h_i)dh_i \Big)^{3/5} ,
\label{correction}
\end{equation}
where $C_n^2(h_i)dh_i$ refers to the turbulence strength of each of each of the six MASS layers, $C_n^2(GL)dh_{\mathrm{GL}}$ is the MASS-DIMM computed GL turbulence strength 
and $f_c(ws,wd)$ is a correction factor, which we take as median correction for each wind speed and direction ($ws, wd$) configuration, directly from Fig.~\ref{ctiot3GLwsdiff}. 
For northern winds, we use the solid curve from Fig.~\ref{ctiot3GLwsdiff} for the corresponding wind speeds. In the case of southern winds, we assume $f_c = 0.9$ 
for all wind speeds. 

In order to test this correction method, we computed the corrective term $f_c$ from the first half of the simultaneous CTIO--T3 data (14,863 samples), similar to what is 
shown in Fig.~\ref{ctiot3GLwsdiff}. This correction was then applied to the second half of the simulataneous data. Table~\ref{ctiocorrecteddimm} shows the results. 
The differences between the statistical distributions are less than 0\arcsecf03 after the correction is applied, demonstrating the use of the concept.  
\section{Turbulence Parameters during 2004--2008}
\label{20042008}
The discussions in the previous sections allow us to investigate now the behavior of the atmosphere above Cerro Tololo between April 1 2004 and December 1 2008.
In order to be able to apply the corrections developed in the previous sections we  
only make use of MASS, DIMM and weather station data when all three data are available within 60~s and 120~s, respectively. Therefore, the following results 
are based on a total of 433,162 samples. 
\subsection{Overall turbulence conditions}
Table~\ref{tololocumltab} shows the overall statistics of the main turbulence parameters; 
the $\epsilon_{\mathrm{DIMM,corr}}$ values represent the statistics of the corrected DIMM measurements. Apart from covering a longer time span, 
these DIMM data are also corrected for the effects desribed in \S~\ref{biassec} and \S~\ref{corrsection}. 
Also, the GL seeing $\epsilon_{\mathrm{GL, corr}}$ was corrected for the effects. 
The isoplanatic angle $\theta_0$ and the MASS seeing given in Tab.~\ref{tololocumltab} were calculated from the turbulence profiles observed by MASS. 
The coherence time statistics are based on the MASS delivered 
coherence time $\tau_{0,\mathrm{MASS}}$ (\citealt{toko02}) and the correction from \cite{travouillon09} was applied: $\tau_{0,\dagger} = \tau_{0,\mathrm{MASS}} / 0.577$. 
$\tau_{0,\dagger}$ represents the free atmosphere coherence time only; it does not take into account any GL contribution. 

\begin{figure}[h]
\begin{center}
\resizebox{1 \textwidth}{!}{\rotatebox{90}{\includegraphics{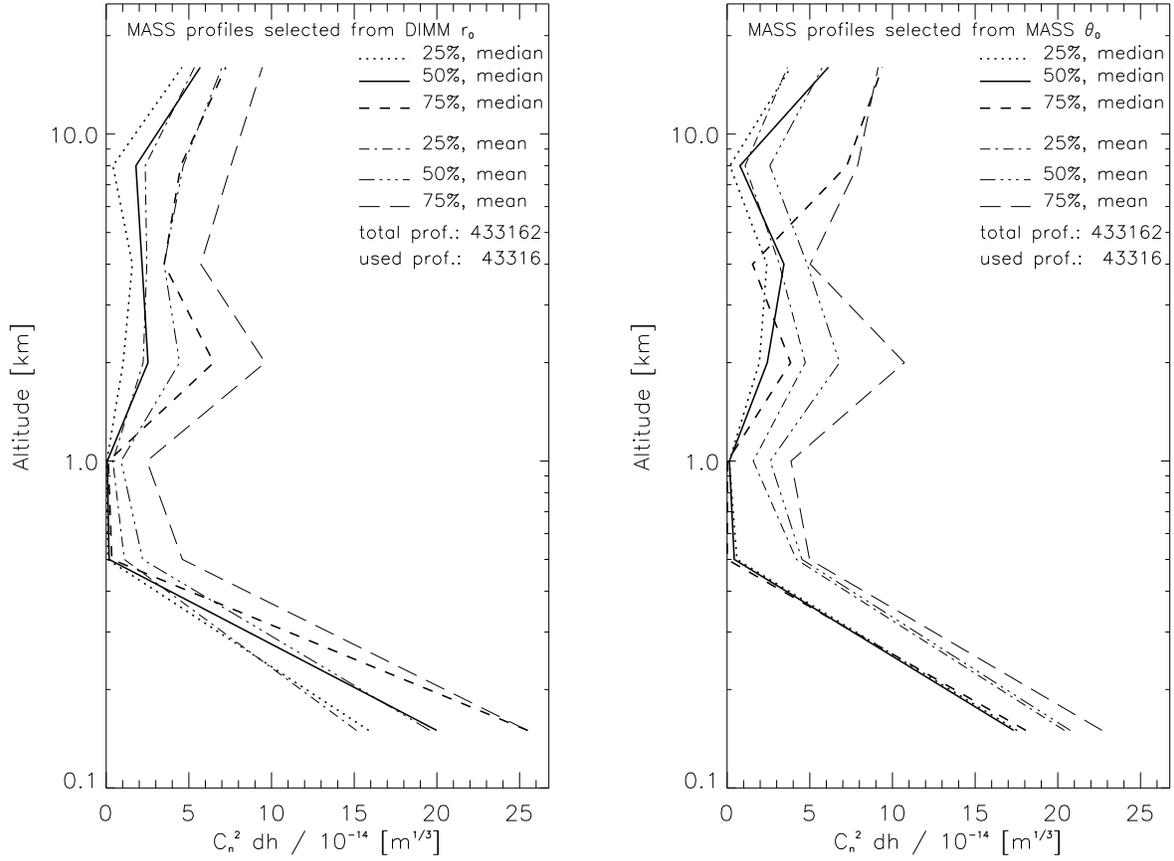}}}
\end{center}
\caption{Turbulence profiles under typical DIMM seeing and isoplanatic angle conditions at Cerro Tololo. 
 }
\label{tololocn2profs}
\end{figure}

The turbulence profiles are shown in Fig.~\ref{tololocn2profs} and detailed in Tab.~\ref{profiltabelle}. These are the profiles under typical seeing and isoplanatic angle 
conditions. They were obtained similar as to what was described in \cite{els09}; by selecting the $\pm$5\% of DIMM $r_0$ (or MASS $\theta_0$) data which result closest in the 
25\%, 50\% and 75\%ile of the overall DIMM seeing (MASS $\theta_0$) statistic. The turbulence profile is then constructed from the MASS data observed simultaneously with the 
selected DIMM $r_0$ (MASS $\theta_0$) data. 

Comparing the total seeing at Tololo to what has been observed at the TMT candidate site in northern Chile, it turns out that the Tololo seeing is by $\approx 0\arcsecf1$ larger. 
The free atmosphere (MASS) seeing is responsible in part for this, as high altitude turbulence is stronger above Tololo by up to 0\arcsecf07. On one side, this reflects itself in the lower 
isoplanatic angle, which is driven by the 16~km layer. But the turbulence profiles in Fig.~\ref{tololocn2profs} also show significant turbulence already at 2~km above Tololo. 
The profiles obtained in northern Chile indicate that significant turbulence strength appears from the 4~km layer onwards. From our findings in \S~\ref{masscomp} we suspect 
that the 2~km layer above Tololo is probably weaker and that its strength is in part an instrumental artifact and that part of this layer's turbulence strength has probably 
to be distributed into the neighboring layers. However, it was shown in \S~\ref{masscomp} that the free atmosphere seeing measured by the TMT and CTIO MASS compare 
extremely well. It can thus be concluded that the free atmosphere at the latitude of Tololo indeed shows slightly more optical turbulence to what 
is found at more northern locations. 
\subsection{Seasonal variations}
\label{seasons}
\begin{figure}[h]
\begin{center}
\resizebox{1 \textwidth}{!}{\rotatebox{90}{\includegraphics{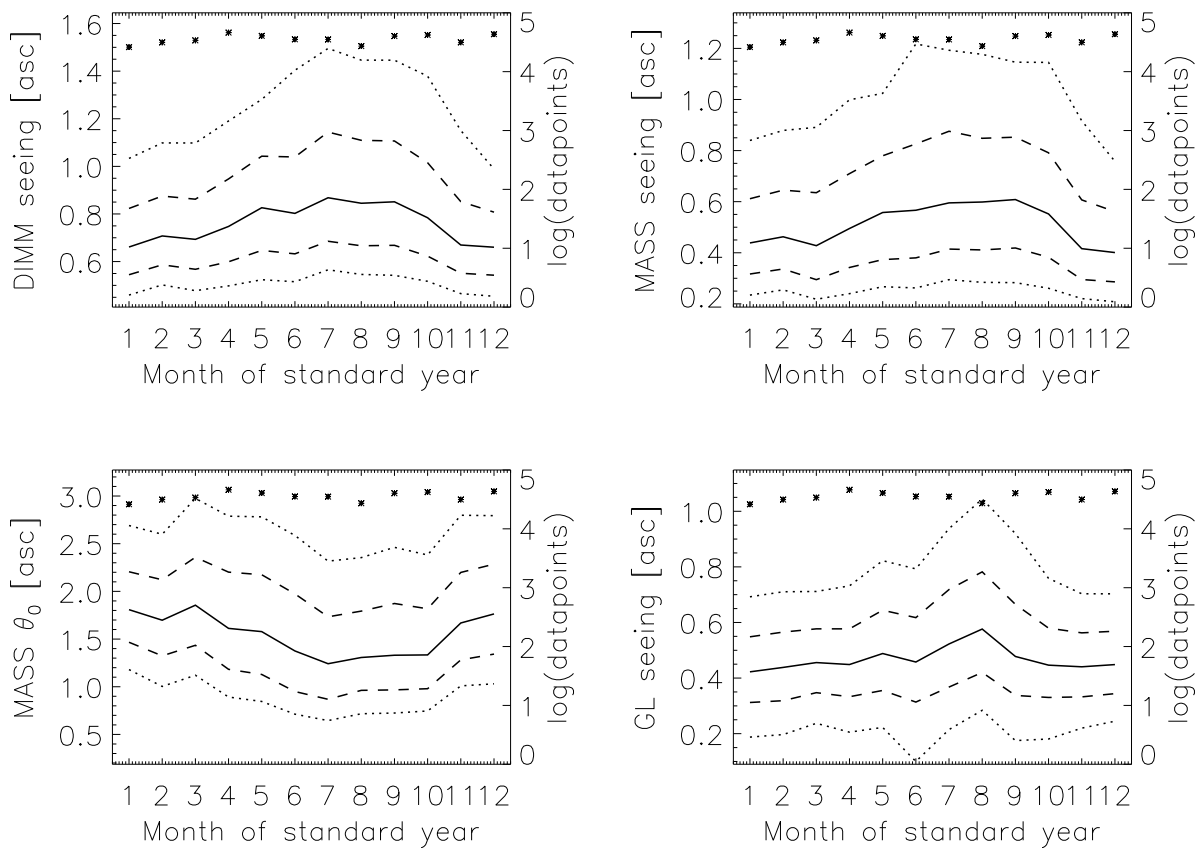}}}
\end{center}
\caption{Main turbulence parameters during the standard year. The dotted lines represent the 10\% and 90\%iels, 
the dashed lines the 25\% and 75\%iles and the solid line the median values, respectively. The dots show the 
amount or data points collected in each month which were used to compute the respective statistics. 
 }
\label{stdyearsee}
\end{figure}

\begin{figure}[h]
\begin{center}
\resizebox{1 \textwidth}{!}{\rotatebox{90}{\includegraphics{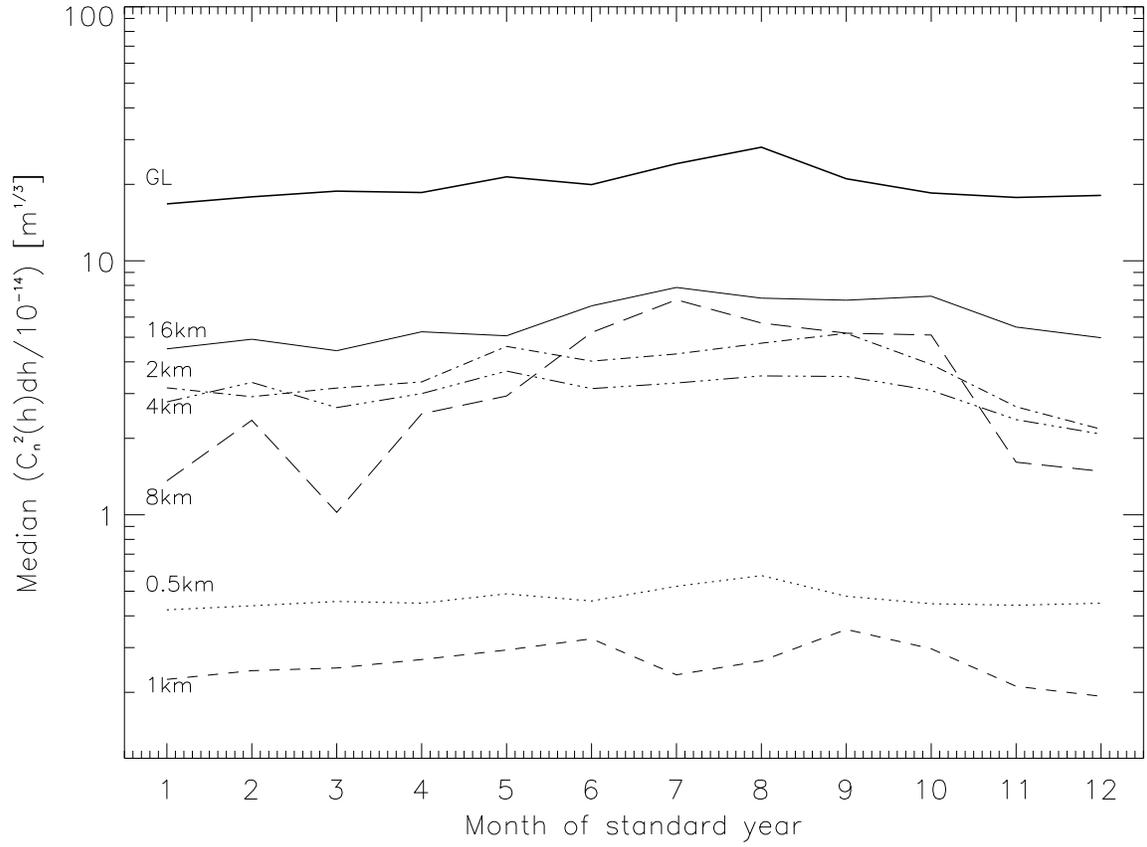}}}
\end{center}
\caption{Median turbulence strength $C_n^2(h)dh$ of the MASS layers during the standard year. 
 }
\label{stdyearcn}
\end{figure}

In order to assess the seasonal variability of the main turbulence parameters we calculate a standard year similar to \cite{els09}. 
It contains the statistics of all data collected in each month of the year covered by the monitoring period. The standard year statistics of MASS, DIMM 
and GL seeing, as well as the isoplanatic angle are shown in the panels of Fig.~\ref{stdyearsee}. The standard year of the median strength of the individual MASS 
layers is shown in Fig.~\ref{stdyearcn}. Both, the MASS and DIMM monthly median seeing components, are stronger during winter time, i.e., from May to October 
by up to 0\arcsecf2. The median GL seeing undergoes a similar annual cycle but on a lower level. 
The amplitude of the seasonal variation of the total seeing at Tololo is comparable to what is found at other sites (\citealt{michel03}, \citealt{masciadri06}, 
\citealt{schoeck09}).
Strong seasonal variation of the isoplanatic angle is observed as well; low during the winter and approximately 0\arcsecf5 higher during 
summer months. The behavior of the individual layer strengths in Fig.~\ref{stdyearcn} shows, that the 8~km and the 16~km layers are 
increased during the winter months. This has already been observed during other site monitoring campaigns in this region (\citealt{vernin00}). 
As these high layers are dominating the isoplanatic angle, their behavior reflects directly onto $\theta_0$. 
The 8~km layer resembles for Tololo the 200~mbar ($\approx$11~km a.s.l.) level and therefore the altitude of the jet stream. The jet stream passes over 
the latitude of Tololo during the southern winter months and it can be expected that it drives the 8~km turbulence strengths. This is very similar to what has been 
observed during the TMT site testing at sites in northern Chile (\citealt{els09}). Also the 1~km, 2~km and 4~km layers might show a similar seasonal variation.
The 0.5~km layer does not inhibit any clear seasonal change; it remains at low levels. This means that the annual cycle of the turbulence strengths of 
the lower tropospheric layers at the latitude of Tololo is inverted as compared to the central Atacama region, i.e., above the northern Atacama, the turbulence strengths 
of the lower layers appear weaker during the winter time. We suspect that the weaker occurrence of the Altiplano Winter (\citealt{zou98}) at Tololo is influencing this 
behavior, but a proper understanding would require additional meteorological data and remains the task for a future study. 
\subsection{Nocturnal variations}
\begin{figure}[h]
\begin{center}
\resizebox{0.7 \textwidth}{!}{\rotatebox{0}{\includegraphics{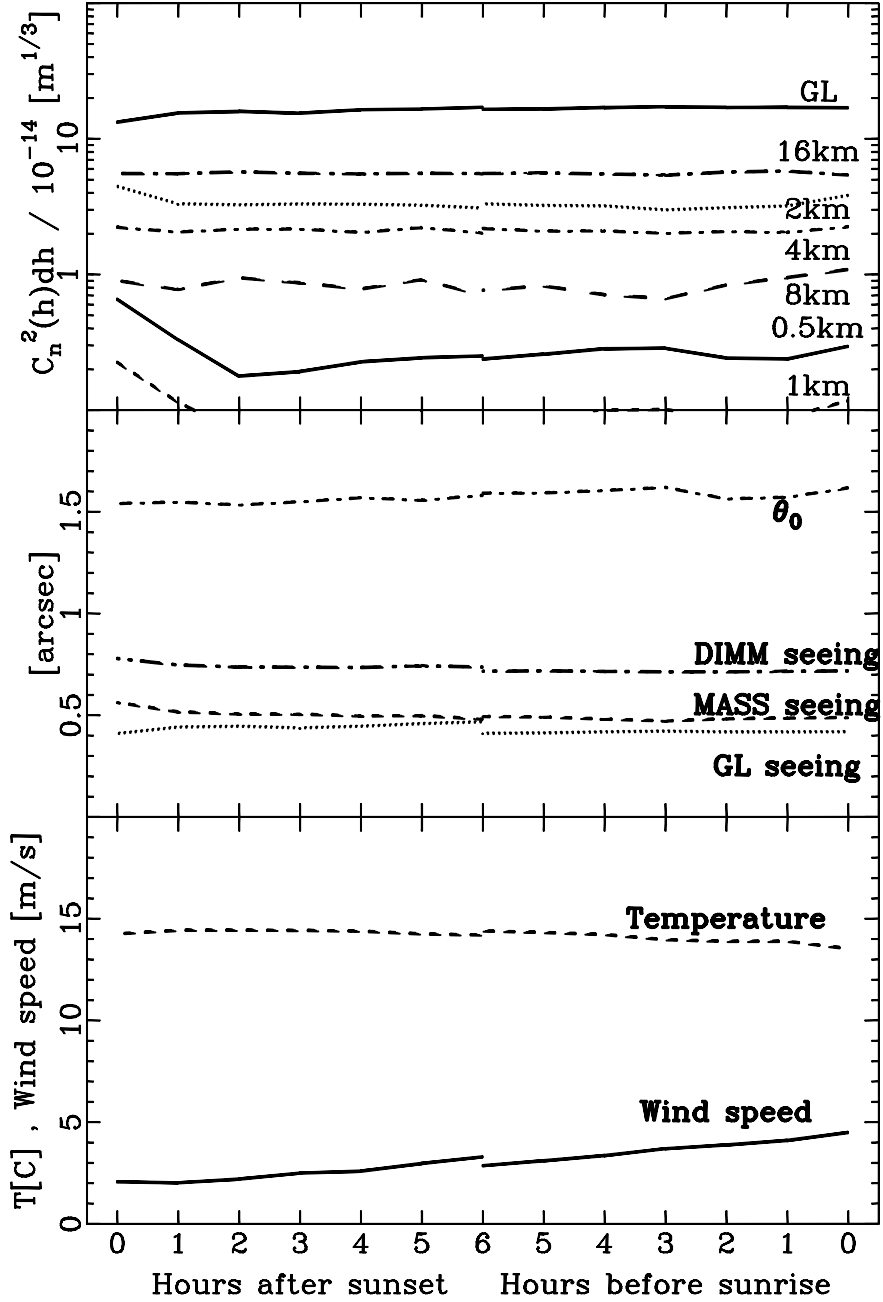}}}
\end{center}
\caption{Median nocturnal variation of atmospheric parameters. 
 }
\label{stdnight}
\end{figure}

The median evolution of the individual MASS layers, the integrated turbulence parameters, wind speed and temperature during a night are shown 
in Fig.~\ref{stdnight}. These graphs show the median values of each parameter for each hour after sunset until 6~hours, and from there on during each hour 
before sunrise. Only during the first two hours of the night do the turbulence strengths (thus seeing) show a significant change. The turbulence strength 
of the GL layer increases, whereas the strength of the layers up to 2~km drop during these hours. This is likely resembling the build up of the stable 
boundary layer and the decay of the inversion layers above. 
Throughout the night the turbulence strengths of the layers remain almost constant, with the 0.5~km layer being the only one showing some small increase during the night. 
This might either reflect the growing vertical extension of the stable boundary layer or as the wind speed is steadily increasing at Tololo, it might also point 
to wind shear as the main mechanism driving the optical turbulence as at the TMT candidate sites. It should be kept in mind, that the particular increase of the GL at Tololo, 
despite having the GL corrected for the dome seeing component, might still be affected by the telescope support structures. 
\section{Summary and Conclusions}
Four years of MASS-DIMM data collected at Cerro Tololo have been presented. These MASS-DIMM data have been compared to simultaneously collected TMT site testing
data during part of this period. This led to the identification of an enhanced ground layer in the CTIO DIMM data under northerly winds. This component is most likely caused 
by the not fully opening dome. A correction of this component has been presented. If this component is indeed due to the dome, this demonstrates that strong seeing biases can be 
introduced within a region extending less than one meter away from the site monitoring telescope. 
The MASS comparison results indicate an exellent agreement of better than 0\arcsecf03 of the measured free atmosphere seeing. The strengths of the individual MASS layers agree well, 
however, individual layers can differ up to $10^{-14}$~m$^{1/3}$. Our simultaneous MASS observations of different stars also confirm the general assumption of uniform layers 
above the site to a level of $0.2\cdot 10^{-14}$m$^{1/3}$.
The corrected DIMM data indicate a median seeing at Cerro Tololo of 0\arcsecf75. These results are consistent with the findings of previously conducted site monitoring campaigns 
in this region, using different methods than employed in our study (\citealt{vernin00}). 
It is also very close to the model by \cite{racine05}, which suggests a seeing of 0\arcsecf74 for Tololo. Our observations would indicate that Tololo shows a seeing only slightly 
larger than what is found in northern Chile. This seeing is in part due to stronger high altitude turbulence. Our data indicate that a strong annual variation 
of higher altitude turbulence exists, with the weakest turbulence encountered during the southern summer months. These results are consistent with the findings of 
previously conducted site monitoring campaigns in this region, using different methods than employed in our study  (\citealt{vernin00}). 
 
After being deployed during three years on Cerro Tolonchar in the North of Chile, the T3 system returned in April 2009 to Cerro Tololo and is now operating within 
$\sim$1~m NNW of its location during the 2004--2005 campaign. As the original CTIO site monitor will continue its operation, this configuration of two MASS-DIMM systems being 
so close to each other, will make it possilbe to conduct a number of experiments. The MASS data could be used to investigate the behavior of optical 
turbulence at various altitudes and over different horizontal spatial scales. In combination with the DIMM, these site monitors could act as an experimental test bed 
for the seeing induced by various dome and shelter configurations, i.e., like a wind tunnel. 
\acknowledgments
We acknowledge the work by CTIO's mountain staff in support of the operation of the CTIO 
site monitor. 
S.G.E. acknowledges discussions with M. Sarazin, J. Thomas-Osip and A. Tokovinin on an 
early version of this manuscript. 
The authors acknowledge the support of the TMT Project and the 
support of the TMT partner institutions. They are the Association 
of Canadian Universities for Research in Astronomy (ACURA), the 
California Institute of Technology and the University of California. 
This work was supported as well by the Gordon and Betty Moore 
Foundation, the Canada Foundation for Innovation, the Ontario 
Ministry of Research and Innovation, the National Research Council 
of Canada, the Natural Sciences and Engineering Research Council 
of Canada, the British Columbia Knowledge Development Fund, the 
Association of Universities for Research in Astronomy (AURA) and 
the U.S. National Science Foundation.

\clearpage
\clearpage
\clearpage
\clearpage
\clearpage
\clearpage
\clearpage
\clearpage
\clearpage
\clearpage
\clearpage
\begin{table}
\begin{center}
\caption{Statistical comparison of the CTIO and T3 MASS systems, calculated from 28,029 simultaneous observations of the same stars. }
\label{ctiot3masscomptab}
\begin{footnotesize}
\begin{tabular}{c c c c c}
\tableline\tableline
MASS result	&	CTIO$_{\mathrm{median}}-$T3$_{\mathrm{median}}$	& CTIO$_{\mathrm{mean}}-$T3$_{\mathrm{mean}}$	&  (CTIO$-$T3)$_{\mathrm{rms}}$	\\
\tableline
seeing [arcsec]							&	 -0.017					&	-0.027					&	0.159				\\
$\theta_0$ [arcsec]						&	 -0.113					&	-0.112					&	0.160				\\
$\tau_0$ [ms]							&	  -0.050				&	-0.006					&	0.295				\\
$C_n(h)^2dh, h=0.5~$km [m$^{1/3}$]  	&	 -4.10$\cdot 10^{-15}$	&	-2.69$\cdot 10^{-14}$	&  3.12$\cdot 10^{-13}$  \\
$C_n(h)^2dh, h=1~$km [m$^{1/3}$]		&	 -1.53$\cdot 10^{-15}$	&	-7.20$\cdot 10^{-15}$	&  9.73$\cdot 10^{-14}$  \\
$C_n(h)^2dh, h=2~$km [m$^{1/3}$]		&	  1.31$\cdot 10^{-14}$	&	 1.49$\cdot 10^{-14}$	&  6.45$\cdot 10^{-14}$  \\
$C_n(h)^2dh, h=4~$km [m$^{1/3}$]		&	 -9.67$\cdot 10^{-15}$	&	-9.92$\cdot 10^{-15}$	&  4.53$\cdot 10^{-14}$  \\
$C_n(h)^2dh, h=8~$km [m$^{1/3}$]		&	 -2.04$\cdot 10^{-14}$	&	-1.08$\cdot 10^{-14}$	&  3.83$\cdot 10^{-14}$  \\
$C_n(h)^2dh, h=16~$km [m$^{1/3}$]		&	  1.24$\cdot 10^{-14}$	&	 1.63$\cdot 10^{-14}$	&  2.74$\cdot 10^{-14}$  \\
\tableline
\end{tabular}
\end{footnotesize}
\end{center}
\end{table}
\clearpage
\begin{table}
\begin{center}
\caption{Effect of the correction for CTIO longitudinal DIMM seeing measurements using eq.~\ref{correction}. The correction 
was calibrated using the first half of the simultaneous CTIO--T3 MASS-DIMM data. The table shows the cumulative statistics of 
the second half sample only, thus comprising 14,863 individual measurements. }
\label{ctiocorrecteddimm}
\begin{tabular}{crrrrrrrrrrr}
\tableline\tableline
Percentile	&	CTIO DIMM	& CTIO DIMM	corr. & T3 DIMM  \\
	 		&	[arcsec]	&	[arcsec]	&	[arcsec] \\
\tableline
10   &	 0.60  	&	0.55    & 0.58  \\
25	 &	 0.72	&	0.65	& 0.66	\\
50	 &	 0.89	&	0.79	& 0.79  \\
75	 &	 1.11	&	0.99	& 0.98  \\
90	 &	 1.39	&	1.23	& 1.20  \\
\tableline
\end{tabular}
\end{center}
\end{table}
\clearpage
\begin{table}
\begin{center}
\caption{ Cumulative statistics of the turbulence parameters of Tololo. $\epsilon_{\mathrm{DIMM, corr}}$ and $\epsilon_{\mathrm{GL, corr}}$ 
have been corrected for the dome seeing contribution according to the procedure outlined in \S~\ref{corrsection}, whereas $\epsilon_{\mathrm{DIMM}}$ has not. 
$\tau_{0,\mathrm{MASS,}\dagger}$ includes the correction factor from \cite{travouillon09}. }
\label{tololocumltab}
\begin{tabular}{c c c c c c c}
\tableline\tableline
Percentile	&   $\epsilon_{\mathrm{DIMM}}$ & $\epsilon_{\mathrm{DIMM, corr}}$ &	$\epsilon_{\mathrm{MASS}}$ & $\epsilon_{\mathrm{GL, corr}}$	& $\theta_{0,\mathrm{MASS}}$  & $\tau_{0,\mathrm{MASS,}\dagger}$ \\
       &  [arcsec] &  [arcsec] & [arcsec] & [arcsec] & [arcsec] &  [ms] \\
\tableline
 10 &  0.60 & 0.50 &  0.25 &  0.19 &  2.66 & 7.08 \\
 25 &  0.71 & 0.60 &  0.34 &  0.33 &  2.11 & 4.85 \\
 50 &  0.88 & 0.75 &  0.50 &  0.44 &  1.56 & 2.94 \\
 75 &  1.10 & 0.97 &  0.72 &  0.58 &  1.12 &  1.77 \\
 90 &  1.39 & 1.27 &  1.03 &  0.76 &  0.82 &  1.15 \\
\tableline
\end{tabular}
\end{center}
\end{table}
\clearpage
\begin{table}
\begin{center}
\caption{ MASS $C_n^2~(h)dh$ turbulence profiles above Tololo. The top part of the table shows the cumulative statistics of all data of 
each layer. The profiles in the central and lower part of the table were constructed from 43,316 individual ones, representing 10\% of all profiles. 
The median/mean are computed from the 10\% around the 25, 50 and 75 percentiles of the DIMM $r_0$ (central part of table) and MASS $\theta_0$ 
(bottom part of table). }
\label{profiltabelle}
{\footnotesize
\begin{tabular}{c  c c c  c c c}
\hline
\hline
         &                 \multicolumn{5}{c}{$C_n^2(h)~dh ~[m^{1/3}]$}        \\
h [km]   &   10\%   &   25\%               &  50\%  &   75\%   &  90\%     \\
\hline
0.0   &      1.055e-13  &       1.830e-13  &       2.876e-13  &       4.224e-13  &       6.119e-13 \\
0.5   &      2.248e-16  &       5.619e-16  &       2.310e-15  &       2.769e-14  &       1.041e-13 \\
1.0   &      1.930e-16  &       4.826e-16  &       9.652e-16  &       9.167e-15   &      6.039e-14 \\
2.0   &      6.028e-16  &       5.694e-15  &       3.146e-14  &       9.569e-14   &      2.246e-13 \\
4.0   &      4.622e-16   &      2.415e-15   &      2.284e-14  &      5.866e-14   &      1.185e-13 \\
8.0   &      3.033e-16  &       7.581e-16   &      1.150e-14  &       6.824e-14  &       1.384e-13 \\
16.0   &      2.166e-14   &      3.408e-14   &      5.661e-14  &       9.514e-14  &       1.605e-13 \\
\hline
\hline
         &                 \multicolumn{3}{c}{median $C_n^2(h)~dh ~[m^{1/3}]$} &  \multicolumn{3}{c}{mean $C_n^2(h)~dh ~[m^{1/3}] $}   \\
h [km]   &   25\%  $r_0$   &   50\% $r_0$               &  75\% $r_0$  &   25\%  $r_0$   &  50\% $r_0$             &   75\% $r_0$ \\
\hline
 0.0 &     1.59e-13 &     2.00e-13 &     2.55e-13 &     1.51e-13 &     1.96e-13 &     2.55e-13 \\
 0.5 &     9.44e-16 &     1.78e-15 &     3.43e-15 &     1.10e-14 &     2.20e-14 &     4.61e-14 \\
 1.0 &     4.34e-16 &     6.80e-16 &     1.29e-15 &     4.01e-15 &     9.02e-15 &     2.46e-14 \\
 2.0 &     1.03e-14 &     2.53e-14 &     6.50e-14 &     2.24e-14 &     4.44e-14 &     9.61e-14 \\
 4.0 &     1.58e-14 &     2.17e-14 &     3.51e-14 &     2.46e-14 &     3.50e-14 &     5.72e-14 \\
 8.0 &     4.07e-15 &     1.80e-14 &     4.51e-14 &     2.36e-14 &     4.66e-14 &     7.54e-14 \\
16.0 &     4.58e-14 &     5.69e-14 &     7.23e-14 &     5.34e-14 &     7.00e-14 &     9.44e-14 \\\hline
\hline
h [km]   &   25\%  $\theta_0$   &   50\% $\theta_0$               &  75\% $\theta_0$  &   25\%  $\theta_0$   &  50\% $\theta_0$             &   75\% $\theta_0$ \\
\hline
 0.0 &     1.75e-13 &     1.73e-13 &     1.81e-13 &     2.04e-13 &     2.08e-13 &     2.27e-13 \\
 0.5 &     5.93e-15 &     4.31e-15 &     2.74e-17 &     4.19e-14 &     4.53e-14 &     4.98e-14 \\
 1.0 &     1.31e-15 &     1.41e-15 &     2.19e-16 &     1.58e-14 &     2.61e-14 &     3.86e-14 \\
 2.0 &     1.95e-14 &     2.43e-14 &     3.84e-14 &     4.73e-14 &     6.80e-14 &     1.07e-13 \\
 4.0 &     2.40e-14 &     3.44e-14 &     1.54e-14 &     3.16e-14 &     4.84e-14 &     5.03e-14 \\
 8.0 &     1.71e-15 &     7.77e-15 &     7.24e-14 &     1.09e-14 &     2.59e-14 &     7.89e-14 \\
16.0 &     3.85e-14 &     6.12e-14 &     9.39e-14 &     3.65e-14 &     5.76e-14 &     9.14e-14 \\
\hline
\end{tabular}
}
\end{center}
\end{table}

\end{document}